\documentclass{PoS}

\pdfoutput=1 

\usepackage[round]{natbib}

\def\farcs{\hbox{$.\!\!^{\prime\prime}$}}

\title{\href{http://www.astro.uni-bonn.de/~wucknitz/publications/pub.php?2008_evn9_1131-1231}{The gravitational lens J1131--1231 ---\\ How to avoid missing an
  opportunity}}

\ShortTitle{The gravitational lens J1131--1231}

\author{\speaker{Olaf Wucknitz}%
  \thanks{This work is supported by the Emmy-Noether-Programme of the
    `Deutsche Forschungsgemeinschaft', reference WU\,588/1-1.}\\
  Argelander-Institut f\"ur Astronomie, Auf dem H\"ugel 71, 53121
  Bonn, Germany\\
  E-mail: \email{wucknitz@astro.uni-bonn.de}}

\author{Filomena Volino\\
  Argelander-Institut f\"ur Astronomie, Auf dem H\"ugel 71, 53121
  Bonn, Germany\\
  E-mail: \email{fvolino@astro.uni-bonn.de}}

\abstract{So far the lens J1131--1231 has been studied only at optical and
  X-ray wavelengths. A detection in the radio was almost missed as a result of
  an incorrect position and archive problems. A direct analysis of NVSS $uv$
  data --- in contrast to the catalogue or images alone --- provided
  sufficient evidence of a detection to justify further radio investigations.
  The system was subsequently observed with MERLIN and the EVN in e-VLBI mode.
  Even though MERLIN seems to show the lensed star-forming regions \emph{and}
  the compact cores, a preliminary analysis of the EVN data only shows an AGN
  in the lens itself but not the \emph{lensed cores.} Additional VLA
  observations will be carried out soon.}

\FullConference{The 9th European VLBI Network Symposium on The role of VLBI in the Golden Age for Radio Astronomy and EVN Users Meeting\\
  September 23-26, 2008\\
  Bologna, Italy}

\begin{document}

\section{Introduction}

The background source in RXS~J1131--1231 consists of an irregular star-forming
galaxy with a Seyfert core at $z=0.658$. While the core is quadruply imaged,
the host galaxy is distorted into an Einstein ring of 3\farcs6 diameter.
J1131--1231 is unrivalled in the amount of detail in the lensed image
configuration, providing a wealth of constraints for lens modelling purposes.
Many of the components are highly magnified so that a detailed study of the
background source becomes possible.  As noted by \citet{sluse03} and
\cite{claeskens06}, the flux ratios of the Seyfert cores deviate considerably
from the model expectations.  Possible explanations are substructure in the
mass distribution, differential dust extinction or microlensing.

\section{Motivation for radio studies}

In contrast to optical wavelengths, radio observations are not impaired by
microlensing and dust extinction, and are therefore essential to resolve flux
anomalies like observed in this system.

\begin{figure}[htb]
\begin{center}
\includegraphics[width=0.25\textwidth,angle=-90]{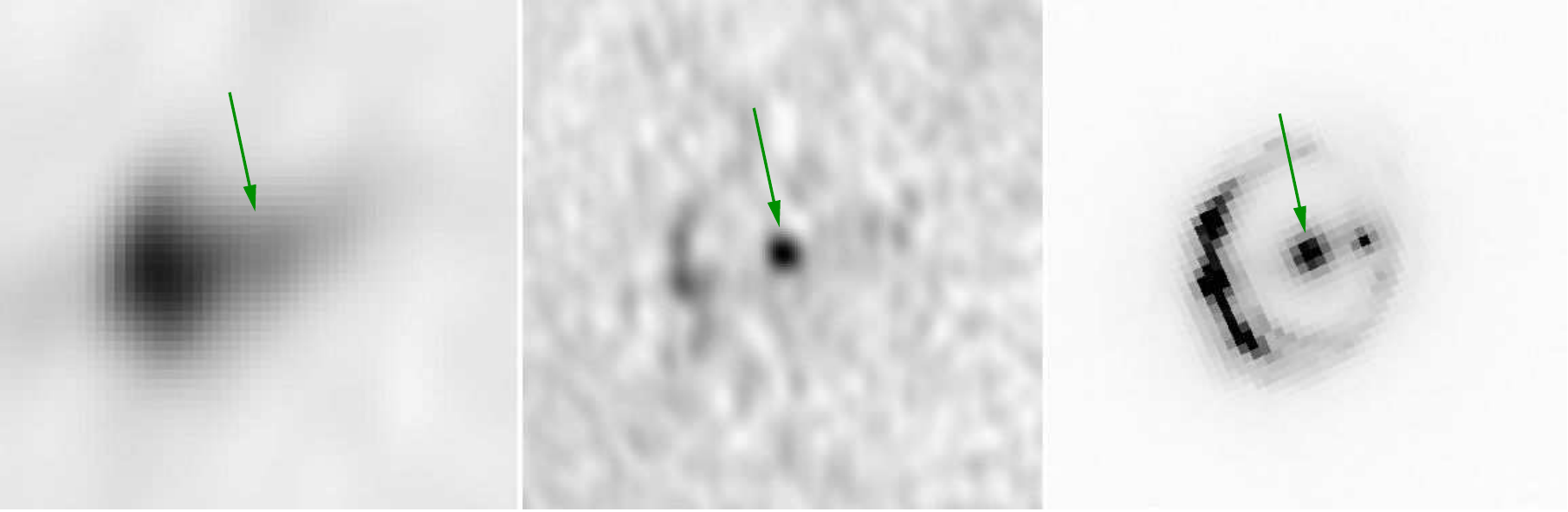}
\end{center}
\caption{VLA snapshot maps in L-band and C-band with an HST image for
  comparison. The arrows mark the position of the lensing galaxy which seems
  to harbour an AGN.}
\label{fig:snapshot}
\end{figure}

The NVSS \citep{condon98} catalogue lists an extended source of 29\,mJy close
to the optical position of the lens. Fitting directly to the NVSS visibility
data, we found that this source actually consists of at least three
sub-components, one of which (6\,mJy) is consistent with being the radio
counterpart of the lens.  Using an archived VLA snapshot, we were later able
to confirm this identification (Fig.~\ref{fig:snapshot}).  The maps show an
AGN (0.75\,mJy) with jets \emph{in the lensing galaxy,} and possible
counterparts of the lensed Seyfert core (0.5--1.3\,mJy each) and star-forming
regions (1--2\,mJy).  The total flux density is consistent with our initial
estimate from the NVSS visibilities.  In addition (not shown here), we see two
radio lobes connected to the AGN in the lensing galaxy.  These lobes
correspond to the other two subcomponents of the NVSS source.

\section{Higher resolution}

\begin{figure}[htb]
\center\includegraphics[width=0.3\textwidth,angle=-90]{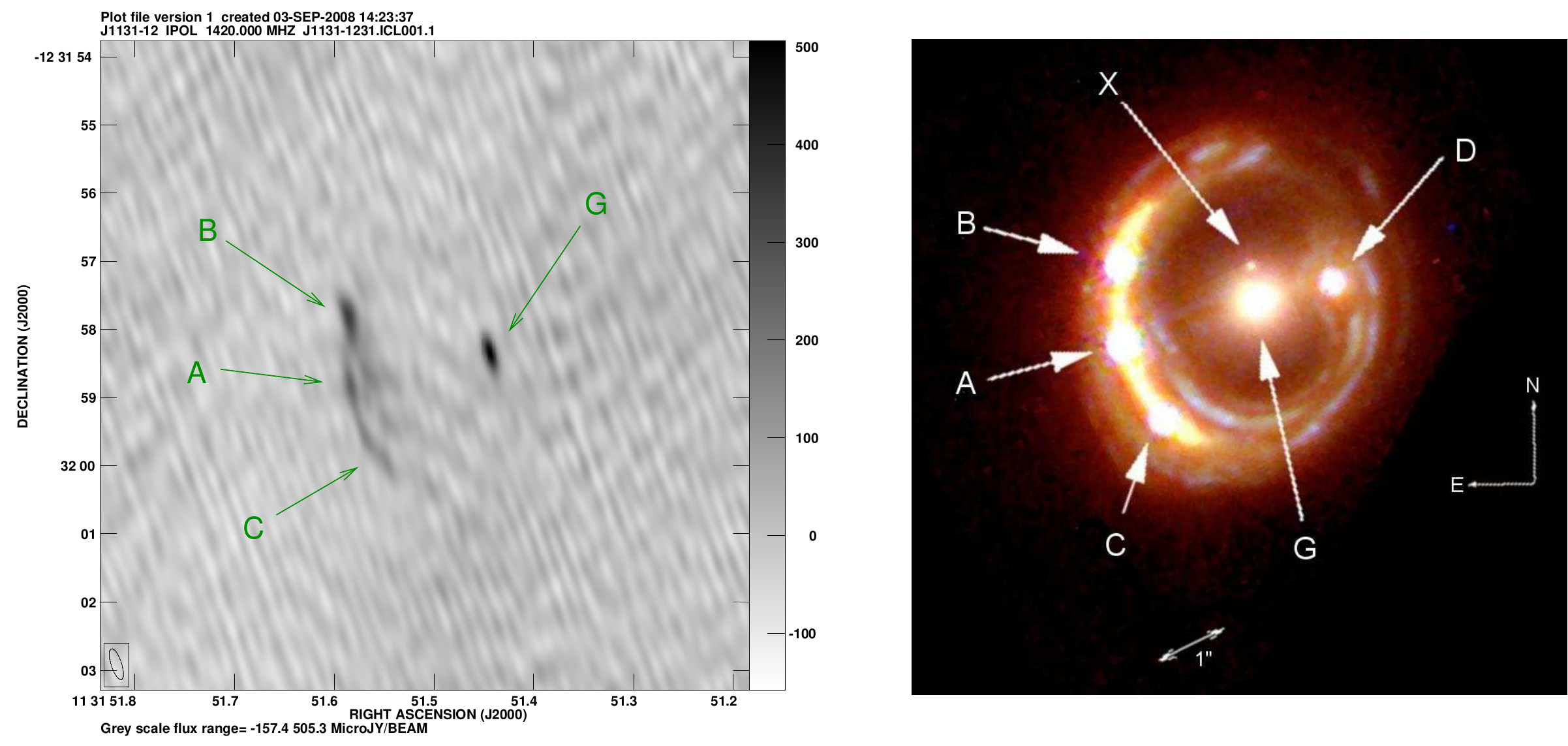}
\caption{Preliminary L-band MERLIN map, 40~h at frequencies 1420 and
  1658\,MHz, compared to the HST image \citep[adapted from][]{claeskens06} at
  the same orientation and scale. The lensing galaxy's core \emph{G} and the
  arc connecting \emph{A,B,C} are clearly visible.  }
\label{fig:merlin}
\end{figure}

In May 2008 we observed the system with MERLIN in L-band.
Fig.~\ref{fig:merlin} shows a preliminary map produced after removing
interfering sources with an own peeling algorithm.  We are inclined not to
trust all the components before our exploration of the reliability is
completed.

Most of the emission detected seems to originate from the star-forming
regions. The compact Seyfert core is weaker than expected from the VLA
observations.

In order to explore the feasibility of detailed VLBI studies, we recently
(June 2008) carried out a short (1\,h on source) e-VLBI experiment at 18\,cm,
using six telescopes (Cm, Mc, Ef, JB1, Tr, Wb) with a data-rate of 512\,Mb/s.
We clearly detected the core of the lensing galaxy at a flux level of
$\sim0.5$--0.7\,mJy but see no hints of the lensed Seyfert core images down to
$<100\,\mu$Jy. This supports the evidence that most of the lensed radio flux
is due to star formation in the background source.

\section{Future projects}

Further VLA time has now been granted to observe this system at C-band in
A-configu\-ration to produce a much deeper version of Fig.~\ref{fig:snapshot}
(centre) and allow a comparison with the L-band MERLIN map.  In addition to
studying the lensed source and the mass distribution of the lens, this system
offers the rare opportunity to see a background source \emph{through} the jet
emanating from the AGN core of the lensing galaxy.  This can potentially be
used to study the physical conditions in the jet.

\section{Conclusions}

J1131--1231 has been detected at radio wavelengths and shows structures of the
lensed star-forming galaxy. It offers the opportunity to study physical
conditions in an AGN jet via propagation effects acting on radiation from the
lensed background source.

Original visibility data of radio surveys are an invaluable source of
information being superior to maps alone. Without the availability of NVSS
visibilities, the information from the survey catalogue or maps alone would
not have been sufficient to motivate further investigations.

\bigskip \emph{If you are conducting a radio survey --- and can afford it ---
  please store the visibilities!}

\bibsep0.ex
\small
\bibliographystyle{aa}
\bibliography{proceedings}

\begin{thebibliography}{3}
\expandafter\ifx\csname natexlab\endcsname\relax\def\natexlab#1{#1}\fi

\bibitem[{{Claeskens} {et~al.}(2006){Claeskens}, {Sluse}, {Riaud}, \&
  {Surdej}}]{claeskens06}
{Claeskens}, J.-F., {Sluse}, D., {Riaud}, P., \& {Surdej}, J. 2006, A\&A, 451,
  865

\bibitem[{{Condon} {et~al.}(1998){Condon}, {Cotton}, {Greisen}, {Yin},
  {Perley}, {Taylor}, \& {Broderick}}]{condon98}
{Condon}, J.~J., {Cotton}, W.~D., {Greisen}, E.~W., {et~al.} 1998, AJ, 115,
  1693

\bibitem[{{Sluse} {et~al.}(2003){Sluse}, {Surdej}, {Claeskens},
  {Hutsem{\'e}kers}, {Jean}, {Courbin}, {Nakos}, {Billeres}, \&
  {Khmil}}]{sluse03}
{Sluse}, D., {Surdej}, J., {Claeskens}, J.-F., {et~al.} 2003, A\&A, 406, L43

\end{thebibliography}

\end{document}